# Quantifying power flow processes mediated by spin currents

Kenta Nakahashi[1], Kohei Takaishi[1], Takayuki Suzuki[2], Katsuichi Kanemoto[1,3]*

1 *Department of Physics, Graduate School of Science, Osaka City University*
*3-3-138 Sugimoto, Sumiyoshi-ku, Osaka 558-8585, Japan.*

2 *JEOL RESONANCE Inc., 3-1-2 Musashino, Akishima, Tokyo 196-8558, Japan.*

3 *Nambu Yoichiro Institute of Theoretical and Experimental Physics (NITEP), Osaka City University*
*3-3-138 Sugimoto, Sumiyoshi-ku, Osaka 558-8585, Japan.*

* To whom correspondence should be addressed; E-mail: kkane@sci.osaka-cu.ac.jp

**Abstract**.
The power flow process mediated by spin current in the bilayer device consisting of ferromagnetic metal (FM) and non-magnetic metal (NM) layers is examined by realizing experimental evaluations for each process from the microwave absorption to electromotive force (EMF) output. The absorption power by ferromagnetic resonance (FMR) of the thin FM layer during the EMF output is directly measured in *operando* using an antenna probe system. The transfer efficiency of the absorption power into the NM layer by spin pumping is estimated from strict linewidth evaluation of EMF spectra. The maximum transfer efficiency of the spin pumping power to the external load via the inverse spin Hall effect is determined to be $4.2\times10^{-8}$ under 160mW microwave irradiation using an analysis model assuming a parallel circuit. The main factors reducing the efficiency are found to be low resistivity of the NM layer and the interface loss. These quantifications are important as a first step to consider the efficient transfer of spin energy mediated by spin currents.

## I. INTRODUCTION

Efficient transfer of spin can be realized by a pure spin current that carries spin angular momentum without charge flow in its parallel direction [1,2]. For practical application of spin current, not only development of spin manipulation and device fabrication techniques but also establishment of effective evaluation methods of spin current propagation is required. One of the most common method employed so far for the spin current evaluation is the combination of spin pumping by ferromagnetic resonance (FMR) for a ferromagnetic metal (FM) layer and observation of electromotive force (EMF) that is induced at a non-magnetic metal (NM) layer by the inverse spin Hall effect (ISHE) via the spin current in an FM/NM bilayer device [3]. The evolution of the observation method for the ISHE process is thus important to advance the understanding of the spin current propagation.

A series of processes from FMR to the EMF output via ISHE can be regarded as transfer processes of spin excitation power. Evaluating the transfer efficiency of the spin power is an interesting topic to the research field of spintronics. The series of processes is classified into the microwave power absorption, the propagation of the absorption power to the NM layer, the conversion of spin current into charge current, and the power output to the external load, each of which has an inherent conversion efficiency. Thus, in order to consider the overall power transfer efficiency, each process has to be evaluated individually. In reality, however, there are some issues with each process evaluation. For example, the conversion process of spin current into charge current has been typically evaluated by the spin Hall angle $\theta_{\text{sh}}$ which is defined by $\boldsymbol{J}_\text{c} = \theta_{\text{sh}}(2e/\hbar)\boldsymbol{J}_\text{s} \times \boldsymbol{\sigma}$, where $\boldsymbol{J}_\text{c}$ and $\boldsymbol{J}_\text{s}$ are the charge current and spin current, respectively, $e$ is the electron charge, $\hbar$ is the reduced Planck constant, and $\boldsymbol{\sigma}$ is the unit vector of the spin polarization. However, since $J_\text{s}$ is not directly measured, $\theta_{\text{sh}}$ usually needs to be estimated by considering a physical model [4-16], which could lead to variation in the estimated $\theta_{\text{sh}}$ values depending on models employed. Furthermore, although EMF per incident microwave power may be regarded as a simple measure of the power transfer efficiency, the absorption power of device may differ greatly even with the same incident power depending on the device size and experimental system employed, and the absorption power itself has not been directly measured. Therefore, the evaluation of the overall power transfer efficiency has not been practically conducted. To achieve this, it is especially desirable to establish a comprehensive experimental evaluation system that can evaluate each process with direct and clear-cut experimental data.

In this paper, we perform comprehensive evaluations of the flow of power absorbed in the FM layer by realizing individual evaluations for each process from the microwave absorption to EMF output via spin currents for the FM/NM bilayer devices. The power flow process is explored in three parts (Fig. 1). First, a new measurement system for microwave absorption using a pick-up antenna is introduced. The antenna system indeed succeeds in measuring the weak absorption power in the thin FM layer. Next, the spin pumping power transferred from the FM layer into the NM layer is estimated by comparison of FMR linewidths between the FM monolayer and the FM/NM bilayer devices, which were designed to minimize linewidth components that are not directly related to the spin pumping. By considering the efficiency of these processes, we realize comprehensive evaluation of the power transfer processes and unravel factors reducing the transferred power. They are particularly performed only from experimental results without a complicated physical model. These analyses can quantitatively evaluate the power transfer processes of the excited spins and thus provide important evaluation guidelines for the development of spin current mediated technology.

## II. EXPERIMENT

The devices fabricated were the permalloy (Py)-monolayer (Py-only) and Py/Pt bilayer devices on a quartz substrate. The quartz substrate was first subjected to ultrasonic cleaning in the order of distilled water, acetone, and 2-propanol for 15 minutes each, and then UV-ozone cleaning for 10 minutes. For the Py/Pt device, a 20 nm-thick Pt layer was formed by vacuum evaporation (13mm×2.5mm) on the cleaned quartz substrate. The Py (Ni 78% and Fe 22%) layers (15nm-thickness) were simultaneously formed by vacuum evaporation on the cleaned substrate for the Py-only device and on the Pt layer for the Py/Pt device (6mm×2.5mm). The two Ag electrode lines to pick up electromotive force (EMF) signals were produced by vacuum evaporation using a shadow mask on the Py surface for the Py-only device and on the Pt surface for the Py/Pt device (20nm) so that the upper and lower electrode lines were parallel to each other. The devices were connected to a thin conducting wire with silver paste and then inserted into a glass cell for EMF measurements and sealed in a nitrogen-filled glove box.

For both devices, the EMF measurements were performed for a device arrangement where the static magnetic field was parallel to the two Ag electrode lines. The EMF signals of the Py-only and Py/Pt devices were measured from the Ag electrode lines, with the surfaces of the Ag lines facing front and back between the poles of the electromagnet for the Py/Pt device and Py-only device, respectively. The FMR and FMR-induced EMF measurements were performed using a partially modified conventional X band ESR spectrometer (JES-FE1XG, JEOL Ltd.) with a rectangular $TE_{102}$ microwave cavity (ES-PCX1, JEOL RESONANCE Inc.). The FMR spectra were measured under magnetic field modulation (80Hz), providing first-derivative FMR spectra. For the EMF measurements where the incident microwave power was less than 1W, the EMF signal was measured by a lock-in technique (SR-830, SRS) synchronized with the microwave amplitude modulation (1.0 kHz). Voltage values by the lock-in measurements were corrected to correspond to the peak-to-peak output values. In the measurements of microwave absorption power, the microwave power emitted from a light guide window on the front side of the rectangular $TE_{102}$ cavity was picked up from an antenna and measured by a detection diode (75KA50, Anritsu) connected with the antenna. The absorption power measurement was performed simultaneously with an EMF measurement. All measurements were performed at room temperature.

## III. RESULTS AND DISCUSSION

This paper addresses the flow of the power induced by FMR for the FM layer via spin currents. For this purpose, processes from FMR excitation to EMF- generation are divided into three parts: microwave absorption, pumping power propagation and EMF generation. The details of the processes and the relation of them to the power output at the external load (Fig. 1) are discussed below.

**A. Microwave absorption**

Here, we propose an experiment to examine the power of microwave absorption in the Py layer. In general, the absorption by FMR is measured using a conventional electron spin resonance (ESR) spectrometer or using a vector network analyzer (VNA). In the former case, the incident microwave is stored in a resonator, and a signal proportional to the power of microwave absorption under the resonance condition is detected by a detector diode, but the incident and absorbed microwave power in the cavity cannot be measured by commercial laboratory equipment. By contrast, the VNA allows estimation of absorption power from calculation of a S-parameter given by the ratio of the incident and output powers, but the method is limited to the case of weak microwave irradiation of about several mW and cannot be applied under strong microwave irradiation as in ISHE measurements typically exceeding 100 mW. In this work, for the direct measurement of microwave power in the ISHE condition, we use a new measurement method that combines an antenna probe system shown in Fig. 2a with a modified ESR spectrometer. The rectangular $TE_{102}$ microwave cavity that we employed has a window on the front side, through which a small amount of the microwave inside the cavity is emitted. The power of emission is proportional to the microwave power inside the cavity, thus the power loss by absorption appears as a small change in the emission power. The emitted microwave is picked up by an antenna probe set near the window and measured by a detection diode connected to the antenna. The output voltage from the diode is proportional to the square route of the power of detected microwave (Fig. 2b).

Figure 2c shows the absorption spectrum of the Py/Pt device obtained actually from the diode via the antenna probe as a voltage signal. The figure also shows the EMF spectrum measured from the Pt electrode simultaneously with the antenna measurement, exhibiting almost the same spectrum lineshape. The averaged baseline and peak height voltages after amplification were $V_0$ =8.79 mV and $\Delta V_p$= 0.16 mV, respectively. From the square route power dependence of the diode signal, the fraction of absorption power to the input power is given by $y_{ab} = (\Delta V_p /V_0)^2$ and calculated as $3.3 \times 10^{-4}$ at the peak position, which corresponds to the optical density $(=-\log_{10}(1- y_{ab}))$ of $1.4 \times 10^{-4}$ for microwave.

The microwave power injected into the cavity in the same experimental condition was directly measured by a power detector located at the input port of the cavity and found to be $P_{in}$ =91 mW. From these results, the power of microwave absorption is obtained by $P_{ab}=y_{ab} P_{in}$ and calculated as 30$\mu$W. The same procedure was performed under different microwave irradiation power. Figure 2d shows that $\Delta V_p /V_0$ is nearly the same, corresponding to a constant $y_{ab}$ for various irradiation power. The result of the constant $y_{ab}$ is reasonable under the relatively low irradiation intensity and demonstrates the suitability of the measurement method. We emphasize that this antenna probe method enables measurements of microwave absorption from a device giving rise to EMF by ISHE, corresponding thus to an *operando* absorption measurement for the working ISHE device.

**B. Estimating spin pumping power**

It has been shown that the magnetization dynamics in ferromagnetic materials are well described by the following phenomenological Landau-Lifshitz-Gilbert (LLG) equation [17,18]

$$\frac{d\mathbf{m}}{dt} = -\gamma \mathbf{m} \times \mathbf{H}_{\text{eff}} + \alpha \mathbf{m} \times \frac{d\mathbf{m}}{dt} \quad , \quad (1)$$

where **m** is the magnetization direction, $\gamma$ is the gyromagnetic ratio, $\mathbf{H}_{\text{eff}}$ is the effective magnetic field consisting of the external and demagnetization fields, and $\alpha$ is the Gilbert damping constant proportional to the FMR linewidth. According to the LLG equation, macroscopic magnetization that absorbs microwave power increases the cone angle of precession motion and acquires magnetization of a component orthogonal to the static magnetic field. In the model of spin pumping [19,20], the orthogonal magnetization component of FM layer transfers magnetization to the FM layer, giving rise to a spin current flowing inside the NM layer.

The power absorbed in the FM layer is thus consumed in the FM layer itself or it is consumed for generating the spin current. Therefore, the efficiency $r_{sp}$ at which the power absorbed by the FM layer $P_{ab}$ is converted into the spin pumping power $P_{sp}$ (defined by the power transferred into the NM layer) in the FM/NM device is calculated as follows:

$$r_{sp} = \frac{P_{sp}}{P_{ab}} = \frac{T_{1b}^{-1} - T_{1m}^{-1}}{T_{1b}^{-1}} = \frac{\Gamma_b - \Gamma_m}{\Gamma_b} \quad . (2)$$

$T_{1b}^{-1}$ is the power consumption (or relaxation) rate of the FM layer in the bilayer device, and $T_{1m}^{-1}$ is the power consumption rate inside the FM layer of the bilayer device and can be regarded as equal to that of the FM monolayer device. $\Gamma_b$ and $\Gamma_m$ are the FMR half-width at half-maximum (HWHM) values of the bilayer and monolayer devices and expected to be proportional to $T_{1b}^{-1}$ and $T_{1m}^{-1}$, respectively, by the same coefficient ($T_{1b}^{-1}/\Gamma_b = T_{1m}^{-1}/\Gamma_m$). Therefore, in order to determine $r_{sp}$, accurate measurements and comparison of the linewidth in the bilayer and monolayer devices are required.

Although linewidth evaluation is possible by FMR measurements with an ESR spectrometer or a VNA, such measurements are usually allowed only at much lower microwave irradiation intensities than that in the ISHE measurements due to signal saturation of microwave detection diodes. Also the linewidth may change depending on the microwave irradiation power. Thus the linewidth in the ISHE measurements should be determined directly from an EMF spectrum in the ISHE measurements. However, it has been shown that the EMF signals of the FM monolayer and the FM/NM bilayer devices normally include not only the ISHE signal but also signals from spin rectification effects such as anisotropic magnetoresistance (AMR) and anisotropic Hall effect (AHE) signals [11,21-25]. In particular, although the ISHE signals give symmetric absorption spectrum of Lorentzian lineshape, AMR and AHE signals consist of the sum of symmetric Lorentzian spectrum and an asymmetric dispersion spectrum with Lorentzian linewidth, meaning that a symmetric spectral component of EMF signals may include the AMR and AHE signals as well as the ISHE signal. Thus, for examining ISHE signals, observed EMF spectra are usually analyzed by assuming that they include both of symmetric and asymmetric spectral components and that the Lorentz linewidths of the two components are the same [11,22-25].

It has been shown that such unwanted signals are essentially caused by in-plane or out-of-plane induction currents generated by FMR in the direction perpendicular to the microwave magnetic field [22-24,26]. We have particularly confirmed that these signals are generated even with a slight deviation from the perpendicular direction to the microwave field. We thus carefully designed devices to reduce the dispersion signal: the boundary lines of the Ag electrodes for picking up EMFs were fabricated so that the upper and lower electrodes were parallel to each other as in Fig. 3a and the device was placed so that it was strictly straight in the vertical direction inside the cavity. The results of EMF spectrum measurements obtained from the Py-only and Py/Pt devices with such device structures are shown in Fig. 3b and were analyzed by the following equation:

$$V_{emf} = V_{sym} \frac{\Gamma^2}{(H-H_0)^2+\Gamma^2} - V_{asym} \frac{2\Gamma(H-H_0)}{(H-H_0)^2+\Gamma^2} \quad (3)$$

where $V_{sym}$ and $V_{asym}$ are the EMF-amplitudes of symmetric and asymmetric components, $H$ is the magnetic field and $H_0$ is the FMR center field, and $\Gamma$ is the HWHM used commonly for the symmetric and asymmetric components and corresponds to $\Gamma_b$ and $\Gamma_m$ for the Py/Pt and Py-only devices, respectively. The analysis results are summarized in Table 1, and importantly, in both spectra, the voltage ratio of the asymmetric component to the symmetric component is less than 1%, indicating that the dispersion signal is extremely reduced in both devices. Figure 3c shows that, when the device is not perfectly arranged vertically in the cavity, the asymmetric component originating from the spin rectification effect actually appears. We note that from the results of fit using Eq. (3), the EMF spectra are found to increase the linewidth due to mixing of the dispersion signals, suggesting that the mixing of the dispersion signals may generate additional relaxation paths. The arrangement of the device inside the cavity is thus important to discuss the linewidth of the devices.

The linewidths of both Py-only and Py/Pt devices in Fig. 3a may have some inhomogeneity. However, since the Py layers of both devices were deposited simultaneously during the fabrication processes, the linewidth inhomogeneity is expected to be the same. The contribution of the inhomogeneity is thus subtracted in the linewidth difference between both samples. From these discussions, the linewidth difference between the two devices can precisely provide the contribution of the power component transferred to the Pt layer by spin pumping. The $r_{sp}$ value calculated from Eq. (2) was determined as 0.090.

**C. EMF generation**

The microwave absorption power of spin transferred to the NM layer by spin pumping gives rise to pure spin currents in the NM layer. The spin currents are then converted into transverse charge currents by ISHE and produce EMF. The power from the charge current is finally consumed by resistance components, one of which can be used as a load resistance in device applications. The load resistance $R_L$ is here substituted by the input resistance of a voltmeter. $V_{emf}^2/R_L$ then corresponds to the power that can be output to the external load. Yet, in this case, the charge currents are also consumed by the resistance of the NM layer itself. Thus, we clarify the relationship between the two resistance components.

Figure 4a presents the results of EMF measurements in the Py/Pt device at the FMR peak magnetic field generated under microwave irradiation of 52 mW injected into the cavity. Under the irradiation power, $V_{emf}$ (=$V_{sym}$) values obtained were -1.02 $\mu$V for $R_L = 55\Omega$ and -2.67$\mu$V for $R_L = 1$ M$\Omega$. These results are considered using a model of the parallel circuit consisting of $R_L$ and the resistance of Pt, $R_{pt}$ (Fig. 4b). Under the model, for the observed $V_{emf}$, the currents of $J_L = V_{emf} / R_L$ and $J_{pt} = V_{emf} / R_{pt}$ are expected to flow in the external load and inside the Pt layer, respectively, and the total charge current is given by $J_c = J_L + J_{pt}$. We found that, when the Pt resistance between the EMF measurement terminals (84.5Ω) is used for $R_{pt}$, $J_c$ for $R_L = 55Ω$ and 1MΩ are calculated as 30.6 nA and 31.6nA, respectively, from the equation of $J_c = V_{emf}(1/R_L + 1/R_{pt})$, indicating that the two $J_c$ values are nearly the same as each other. Note that the agreement of $J_c$ for the change of $R_L$ was also confirmed under much stronger pulse microwave irradiation for the same Py/Pt device [27]. These results demonstrate the validity of the parallel circuit model and show that for a change of $R_L$, the total current $J_c$ only branches to $J_{pt}$ and $J_L$ while keeping $J_c$ constant.

Using the above model, we examine how much power generated by spin pumping can be output via spin currents. The characteristics of output current and voltage is obtained by $V_{emf}$-measurements as a function of the resistance of the external terminal $R_L$. According to the model described above, the total current $J_c$ is regarded as constant. Thus, the relation of output current $J_L$ and voltage $V_{emf}$ is obtained by

$$J_L = V_{emf}/R_L = J_c - J_{pt} = J_c - V_{emf}/R_{pt}. \quad (4)$$

The output power $P_{out}$ is then calculated as:

$$P_{out} = J_L V_{emf} = V_{emf}\left(J_c - \frac{V_{emf}}{R_{pt}}\right). \quad (5)$$

$P_{out}$ thus gives a maximum value $P_{out}^{max} = (1/4)J_c^2 R_{pt}$ when $V_{emf} = (1/2)J_c R_{pt}$. In the condition, the relation $J_L = J_{pt}$ is obtained, which corresponds to $R_L = R_{pt}$, meaning that the output power becomes maximum when the external resistance is matched with the resistance of the NM layer.

By a similar procedure, we can consider the $R_L$-dependence of the total power $P_{tot} = J_c V_{emf}$ stored by ISHE for generated $J_c$ that consists of the sum of $P_{out}$ and the power consumed in the Pt-layer. Since $J_c$ is constant while changing $R_L$, $P_{tot}$ changes depending only on $V_{emf}$ and has the maximum of $P_{tot}^{max} = J_c^2 R_{pt}$ in the condition of $R_L = \infty$ where $V_{emf} = J_c R_{pt}$. In this condition, all the power is consumed inside the Pt layer and cannot be output into the external load.

**D. Comprehensive evaluations of power flow processes**

The major goal of spintronics is to propagate spin energy and harness it for applications. In order to realize that, it is essential to accurately understand how efficiently input energy is propagated and where energy loss occurs. In this section, by integrating the results of the previous sections, we directly determine the efficiency of spin energy propagation and chart the dissipation paths of the input energy.

It has been shown that under the irradiation power of $P_{in} = 162$mW, $P_{ab} = y_{ab}P_{in}$ (=54μW) is absorbed, and part of the absorption power (ratio $r_{sp}$ (=0.090) estimated from Eq. (2)) flows out into the Pt layer as the spin pumping power $P_{sp}$. Thus the conversion efficiency of the injected spin pumping power to the maximum output power ($\eta_{out}$) is given by the following equation:

$$\eta_{out} = P_{out}^{max}/P_{sp} = J_c^2 R_{pt}/(4y_{ab}P_{in}r_{sp}) = V_\infty^2/(4y_{ab}P_{in}r_{sp}R_{pt}) \quad (6)$$

where $V_\infty$ is the EMF in the condition of $R_L = \infty$. Using $V_{sym}$ in Table 1 for $V_\infty$, we obtain $\eta_{out} = 4.2 \times 10^{-8}$ for $P_{in} = 162$mW. From this relation, the conversion efficiency of the absorption power to the output power can also be calculated by the relation $\eta_{out,ab} = r_{sp}\eta_{out}$ and is determined as $3.8 \times 10^{-9}$. Moreover, the conversion efficiency of the pumping power to the total power $\eta_{tot}$ consisting of the sum of $P_{out}$ and the power consumed in the Pt-layer power is given by the relation $\eta_{tot} = P_{tot}^{max}/P_{sp} = 4\eta_{out}$. The defined parameters are summarized in Table 2 and the quantitative power flow from excitation to the output determined in this research is summarized by a flow chart in Fig. 5a. We emphasize that these values were not based on unproven assumptions or models, but were determined directly only from experimentally measured values.

As shown above, the determined power conversion efficiencies were all relatively small. Including the reason why the efficiencies are small, we discuss the factors that determine the magnitude of the efficiencies. For the injected spin pumping power $P_{sp}$, part of $P_{sp}$ is converted into spin currents $J_s$, and $J_s$ are converted into charge currents, which then give rise to EMF. These processes are illustrated in Fig. 5b and we here define $J_{c0}$ as the charge current that gives rise to EMF. In the actual power consumption process, $J_c$ (= $J_{pt}$+ $J_L$) flows in the opposite direction to $J_{c0}$ (Fig. 5c). It should be noted here that, defining $V_{emf0}$ as the virtual EMF when there is no power consumption, the observed EMF $V_{emf}$ is given as the resulting EMF while the power generation and consumption are occurring simultaneously and must be much smaller than $V_{emf0}$ due to the small resistance value of $R_{pt}$. This consideration explains the reason that higher EMFs confirmed in metal NM layers are nearly restricted to metals with relatively high resistivity [28,29]. Since the relation that $V_{emf}$ is much smaller than $V_{emf0}$ means that most of charges supplied by $J_{c0}$ is consumed by $J_c$, it follows that the magnitude of $J_c$ is nearly equal to or slightly smaller than that of $J_{c0}$.

From the above model, we can consider the reason that the determined power conversion efficiencies were small. First, since the measured $V_{emf}$ is much smaller than $V_{emf0}$, the determined $P_{tot}^{max}$ and $P_{out}^{max}$ proportional to $V_{emf}$ should be much smaller than the actual power generated from $J_{c0}$ ($V_{emf0}J_{c0}$). Hence the small resistance of Pt layer reducing $V_{emf}$ is one of the major factors of the small conversion efficiencies. Particularly, since the output power $P_{out}$ is proportional to $J_c^2R_{pt}$, adopting a material with larger resistance for the NM layer has a synergistic effect to enhance $P_{out}$. Also, although the transfer efficiency of spin pumping power to the NM layer ($r_{sp}$) was only about 9% in the Py/Py device, this value is expected to be much larger for the FM layer of yttrium iron garnet (YIG), because the FMR linewidth of YIG layer is known to greatly increase by the presence of an adjacent NM layer [30]. Moreover, the spin Hall angle $\theta_{sh}$ is important as the factor to reduce the power conversion efficiencies, which is discussed in a later paragraph.

Apart from the power resulting from the spin pumping, $P_{ab}$ (1- $r_{sp}$) is not used as the spin pumping power and is consumed in the Py layer. The Py layer also induces EMF by itself as shown in the results of the Py-only device. It has been actually reported that the FM mono layer device exhibits FMR-induced EMF [25,31], and such "self" -EMF effects with symmetric EMF spectra have been explained by self-induced ISHE [25] and magnonic charge pumping [31]. We actually observed a symmetric spectrum from the Py-only device as in Fig. 3b. The power consumed in the Py layer of the Py/Pt device should thus be partly converted to the self-EMF. The power of the self-EMF effect in the Py/Pt device is estimated from the EMF of the Py-only device. Using the resistance measured from the Py-terminals ($R_{py}$=31.2Ω) and $V_{sym}$ of the Py-only device in Table 1, the maximum output power of the self-EMF effect $V_{sym}^2/(4R_{py})$ is calculated to be 5.9×10$^{-14}$ W under $P_{in}$ =162mW. This power is 0.30 of $P_{out}^{max}$ of the Py/Pt device ($(1/4)J_c^2 R_{pt}$ =2.0×10$^{-13}$ W)). Also the conversion efficiency of $P_{ab}$ (1- $r_{sp}$) into the maximum self-EMF power is determined as $\eta_{ab}$ =1.2×10$^{-9}$. This efficiency is also shown in Fig. 5a.

$\theta_{sh}$ has often been used as a parameter for discussing the conversion process from spin currents to charge currents [29]. We here introduce for comparison a $\theta_{sh}$-equivalent parameter that can be determined by our experiments. We focus on the flow of electric power to consider the parameter. It was explained above that, in the case of $R_L$=1MΩ, all of the $J_c$ flows inside the Pt layer and the power of $P_{tot}^{max}= J_c^2 R_{pt}$ is consumed at the Pt-resistance. Thus, in the same $R_L$=1MΩ condition, if all the pumping power $P_{sp}$ is converted to $J_c$ without loss in the process of converting $J_s$ to $J_c$ (corresponding to $\theta_{sh}$ =1), $J_c$ reaches $(P_{sp}/R_{pt})^{1/2}$. Considering that $J_c$ obtained actually was $(P_{tot}^{max}/R_{pt})^{1/2}$, we here define the $\theta_{sh}$-equivalent parameter from the ratio of $J_c$ as

$$\theta'_{sh} = \left(\frac{P_{tot}^{max}}{P_{sp}}\right)^{1/2} = \left(\frac{P_{tot}^{max}}{y_{ab} r_{sp} P_{in}}\right)^{1/2}. \quad (7)$$

This parameter corresponds to the spin Hall angle when all the spin pumping power is converted to $J_s$. Since the pumping power may only be partly converted to $J_s$, $\theta'_{sh}$ is smaller than $\theta_{sh}$ ($\theta'_{sh} \leq \theta_{sh}$). Using $P_{tot}^{max}=V_\infty^2/R_{pt}$, we obtain $\theta'_{sh}$=4.1×10$^{-4}$ for $P_{in}$= 162mW. $\theta_{sh}$ has often been estimated for the Py/Pt device and the values widely range from 4×10$^{-3}$ to 0.11 [6-10,12,15,16]. Our value is smaller than the $\theta_{sh}$ values, which could be due to the difference of the definition mentioned above. However, since a spin current is not a quantity measured directly from experiments, the $\theta_{sh}$ values have been determined from analytical calculations by estimating the spin diffusion length from the film thickness dependence of EMF and the mixing conductance using a physical model. In contrast, the $\theta'_{sh}$ value has the advantage of being a parameter obtained directly only from the experiments.

From the definition above, we obtain the relation $\theta'^{2}_{sh}=\eta_{tot} =4\eta_{out}$. The spin-Hall angle is thus important for the power output because it works on $\eta_{out}$ as a square. We here reconsider factors reducing the power conversion efficiencies. It was recently suggested for the Py/Pt device that the spin pumping energy is largely reduced at the interface between the FM and NM layers [16] followed by theoretical studies [32,33]. Since all the spin pumping power may not necessarily contribute to generating $J_s$, such interface power loss is likely to occur in the present device. In the model described above, when there is the interface loss, it reduces $\theta'_{sh}$ but appears more directly as the reduction in $J_c$, meaning that the magnitude of $J_c$ includes the information of the interface loss. Also, as described above, the power loss due to the low resistance appears in $V_{emf}$. Therefore, the two factors reducing the conversion efficiencies can be evaluated from $J_c$ and $V_{emf}$ and the contributions from them can be discriminated.

The power conversion efficiencies estimated in this study were relatively small. However, the major factors reducing them can be discriminated by evaluating $J_c$ and $V_{emf}$, and optimizing these factors through systematic studies using different materials can contribute to enhancing the efficiencies.

We emphasize that the efficiency measurements proposed in this study are also applicable for an EMF caused by FMR-induced heating due to the spin Seebeck effect [34,35], since both the absorption power and output power can be measured even if the heating process is included. Moreover, when the spin pumping power is injected into the NM layer via temperature gradient due to the spin Seebeck effect, the efficiencies such as $\eta_{tot}$ and $\eta_{out}$ are the same as the cases in this study. Therefore, the evaluation methods proposed in this study are widely applicable and can contribute to examining the spin power conversion and transfer processes of various spintronics systems.

## IV. CONCLUSION

The power flow process mediated by spin current was comprehensively evaluated for the Py/Pt bilayer device by realizing individual evaluations from the microwave absorption to EMF output. The absorption power by FMR in the thin Py layer was directly measured using an antenna probe system. The transfer efficiency of the absorption power into the Pt layer by spin pumping was determined to be about 9% from linewidth comparison of the EMF spectra between the bilayer device and the Py-only device. The total charge current consisting of the currents flowing in the Py layer and the external resistance was shown to be nearly the same when changing $R_L$, which allowed us to assume the total charge current to be constant. The maximum transfer efficiency of the spin pumping power into the external resistance was determined to be $4.2\times10^{-8}$ under 160mW microwave irradiation. We found that the efficiency are reduced primarily due to low resistivity of the NM layer and the interface loss. All the efficiencies in individual processes were estimated from experimental results without complicated models. Evaluation from these procedures can contribute to designing efficient spin current based devices.

## ACKNOWLEDGMENT

This work was supported in part by JSPS KAKENHI Grant (numbers 17H03135 and 19K22172) and by the Osaka City University (OCU) Strategic Research Grant 2019 for basic researches.

## REFERENCES

1. S. Murakami, N. Nagaosa, S. C. Zhang, Dissipationless quantum spin current at room temperature. *Science* **301**, 1348-1351, doi:10.1126/science.1087128 (2003).
2. S. A. Wolf, D. D. Awschalom, R. A. Buhrman, J. M. Daughton, S. von Molnar, M. L. Roukes, A. Y. Chtchelkanova, D. M. Treger, Spintronics: A spin-based electronics vision for the future. *Science* **294**, 1488-1495 (2001).
3. E. Saitoh, M. Ueda, H. Miyajima, G. Tatara, Conversion of spin current into charge current at room temperature: Inverse spin-Hall effect. *Appl. Phys. Lett.* **88**, 182509 (2006).

4. S. O. Valenzuela, M. Tinkham, M. Direct electronic measurement of the spin Hall effect. *Nature* **442**, 176-179 (2006).

5. S. O. Valenzuela, M. Tinkham, M. Electrical detection of spin currents: The spin-current induced Hall effect (invited). *J. Appl. Phys.* **101**, 09B103 (2007).

6. L. Vila, T. Kimura, Y. Otani, Evolution of the spin hall effect in Pt nanowires: Size and temperature effects. *Phys. Rev. Lett.* **99**, 226604 (2007).

7. T. Kimura, Y. Otani, T. Sato, S. Takahashi, S. Maekawa, Room-temperature reversible spin Hall effect. *Phys. Rev. Lett.* **98**, 156601 (2007).

8. K. Ando, S. Takahashi, K. Harii, K. Sasage, J. Ieda, S. Maekawa, E. Saitoh, Electric manipulation of spin relaxation using the spin Hall effect. *Phys. Rev. Lett.* **101**, 036601 (2008).

9. T. Seki, Y. Hasegawa, S. Mitani, S. Takahashi, H. Imamura, S. Maekawa, J. Nitta, K. Takanashi, Giant spin Hall effect in perpendicularly spin-polarized FePt/Au devices. *Nat. Mater.* **7**, 125-129 (2008).

10. M. Morota, K. Ohnishi, T. Kimura, Y. Otani, *J. Appl. Phys.* **105**, 07C712 (2009).

11. K. Ando, E. Saitoh, Inverse spin-Hall effect in palladium at room temperature. *J. Appl. Phys.* **108**, 113925 (2010).

12. O. Mosendz, V. Vlaminck, J. E. Pearson, F. Y. Fradin, G. E. W. Bauer, S. D. Bader, A. Hoffmann, Detection and quantification of inverse spin Hall effect from spin pumping in permalloy/normal metal bilayers. *Phys. Rev. B* **82**, 214403 (2010).

13. M. Morota, Y. Niimi, K. Ohnishi, D. H. Wei, T. Tanaka, H. Kontani, T. Kimura, Y. Otani, Indication of intrinsic spin Hall effect in 4d and 5d transition metals. *Phys. Rev. B* **83**, 174405 (2011).

14. Y. Niimi, M. Morota, D. H. Wei, C. Deranlot, M. Basletic, A. Hamzic, A. Fert, Y. Otani, Extrinsic Spin Hall Effect Induced by Iridium Impurities in Copper. *Phys. Rev. Lett.* **106**, 126601 (2011).

15. L. Liu, T. Moriyama, D. C. Ralph, R. A. Buhrman, Spin-Torque Ferromagnetic Resonance Induced by the Spin Hall Effect. *Phys. Rev. Lett.* **106**, 036601 (2011).

16. X. D. Tao, Q. Liu, B. F. Miao, R. Yu, Z. Feng, L. Sun, B. You, J. Du, K. Chen, S. F. Zhang, L. Zhang, Z. Yuan, D. Wu, H. F. Ding, Self-consistent determination of spin Hall angle and spin diffusion length in Pt and Pd: The role of the interface spin loss. *Sci. Adv.* **4**, eaat1670 (2018).

17. T. L. Gilbert and J. M. Kelly, Anomalous rotational damping in ferromagnetic sheets," in Conf. Magnetism and Magnetic Materials, Pitts- burgh, PA, June 14–16, 1955. NewYork: American Institute of Electrical Engineers, Oct. 1955, pp. 253–263.

18. L. D. Landau, E. M. Lifshitz, L. P. Pitaevski, *Statistical Physics*, Part 2 (Pergamon, Oxford, 1980), 3rd ed.

19. Y. Tserkovnyak, A. Brataas, G. E. W. Bauer, Enhanced Gilber damping in thin ferromagnetic films. *Phys. Rev. Lett.* **88**, 117601 (2002).

20. S. Mizukami, Y. Ando, T. Miyazaki, Effect of spin diffusion on Gilbert damping for a very thin permalloy layer in Cu/permalloy/Cu/Pt films. *Phys. Rev. B* **66**, 104413 (2002).

21. Y. S. Gui, N. Mecking, X. Zhou, G. Williams, C. M. Hu, Realization of a room-temperature spin

dynamo: The spin rectification effect. *Phys. Rev. Lett.* **98**, 107602 (2007).

22. N. Mecking, Y. S. Gui, C. M. Hu, Microwave photovoltage and photoresistance effects in ferromagnetic microstrips. *Phys. Rev. B* **76**, 224430 (2007).

23. A. Azevedo, L. H. Vilela-Leao, R. L. Rodriguez-Suarez, A. F. L. Santos, S. M. Rezende, Spin pumping and anisotropic magnetoresistance voltages in magnetic bilayers: Theory and experiment. *Phys. Rev. B* **83**, 144402 (2011).

24. Z. Feng, J. Hu, L. Sun, B. You, D. Wu, J. Du, W. Zhang, A. Hu, Y. Yang, D. M. Tang, B. S. Zhang, H. F. Ding, Spin Hall angle quantification from spin pumping and microwave photoresistance. *Phys. Rev. B* **85**, 214423 (2012).

25. R. Iguchi, E. Saitoh, Measurement of Spin Pumping Voltage Separated from Extrinsic Microwave Effects. *J. Phys. Soc. Jpn.* **86**, 011003 (2017).

26. A. Tsukahara, Y. Ando, Y. Kitamura, H. Emoto, E. Shikoh, M. P. Delmo, T. Shinjo, M. Shiraishi, Self-induced inverse spin Hall effect in permalloy at room temperature. *Phys. Rev. B* **89**, 235317 (2014).

27. K. Takaishi, K. Nakahashi, T. Suzuki, K. Kanemoto, published elsewhere.

28. H. L. Wang, C. H. Du, Y. Pu, R. Adur, P. C. Hammel, F. Y. Yang, Scaling of Spin Hall Angle in 3d, 4d, and 5d Metals from Y3Fe5O12/Metal Spin Pumping. *Phys. Rev. Lett.* **112**, 197201 (2014).

29. J. Sinova, S. O. Valenzuela, J. Wunderlich, C. H. Back, T. Jungwirth, Spin Hall effects. *Rev. Modern Phys.* **87**, 1213-1259 (2015).

30. M. M. Qaid, T. Richter, A. Muller, C. Hauser, C. Ballani, G. Schmidt, Radiation damping in ferromagnetic resonance induced by a conducting spin sink. *Phys. Rev. B* **96**, 184405 (2017).

31. A. Azevedo, R. O. Cunha, F. Estrada, O. A. Santos, J. B. S. Mendes, L. H. Vilela-Leao, R. L. Rodriguez-Suarez, S. M. Rezende, Electrical detection of ferromagnetic resonance in single layers of permalloy: Evidence of magnonic charge pumping. *Phys. Rev. B* **92**, 024402 (2015).

32. K. Chen, S. F. Zhang, Spin Pumping in the Presence of Spin-Orbit Coupling. *Phys. Rev. Lett.* **114**, 126602 (2015).

33. K. Chen, S. F. Zhang, Spin Pumping Induced Electric Voltage. *IEEE Magn. Lett.* **6**, 2427120 (2015).

34. K. Uchida, S. Takahashi, K. Harii, J. Ieda, W. Koshibae, K. Ando, S. Maekawa, E. Saitoh, Observation of the spin Seebeck effect. *Nature* **455**, 778-781 (2008).

35. K. Uchida, H. Adachi, T. An, T. Ota, M. Toda, B. Hillebrands, S. Maekawa, E. Saitoh, Long-range spin Seebeck effect and acoustic spin pumping. *Nat. Mater.* **10**, 737-741 (2011).

**Table 1** Results of fits using equation (3) for the electromotive force (EMF) spectra obtained under 162mW of microwave irradiation.

| Devices | $V_{sym}$ ($\mu$V) | $V_{asym}$ ($\mu$V) | $V_{asym}/V_{sym}$ (%) | $\Gamma$ (mT) |
|---|---|---|---|---|
| **Py/Pt** | -8.274±0.008 | 0.0046±0.0043 | 0.06 | 2.890±0.004 |
| **Py only** | 2.721±0.003 | 0.011±0.002 | 0.40 | 2.629±0.005 |

**Table 2** Parameters used in this research and their definition.

| Parameter | Physical definition and property | Parameter | Physical definition and property |
|---|---|---|---|
| $P_{in}$ | Incident microwave power loaded into the cavity. | $P_{tot}$ | Total power consisting of the sum of $P_{out}$ and the power consumed in the Pt-layer. |
| $y_{ab}$ | Fraction of absorption power to the input power | $P_{tot}^{max}$ | Maximum output power of $P_{tot}$. |
| $P_{ab}$ | $P_{ab} = y_{ab}P_{in}$ : Power of microwave absorption | $V_{emf}$ | Voltage of electromotive force (emf) |
| $r_{sp}$ | Efficiency at which the power absorbed by the FM layer is converted into the spin pumping power | $V_{sym}, V_{asym}$ | Symmetric and asymmetric emf components used in Eq. (3) |
| $P_{sp}$ | $P_{sp} = r_{sp}P_{ab}$ : Power of spin pumping. | $V_\infty$ | Obtained emf voltage with $R_L$ = 1 MΩ or infinity. |
| $J_s$ | Spin current that flows in the NM layer. | FF | Fill factor of current-voltage characteristics |
| $J_c$ | Charge current that is converted from spin current. | $\eta_{out}$ | Conversion efficiency of the injected spin pumping power to the maximum output power |
| $R_{Pt}$ | Resistance value of NM (Pt) layer. 84.5 Ω. | $\eta_{tot}$ | Conversion efficiency of the injected spin pumping power to the maximum total power |
| $R_L$ | Resistance value of an external load. | $\theta_{SH}$ | Original definition of spin Hall angle. |
| $J_{Pt}$ | Charge current inside the NM (Pt) layer. | $\theta'_{SH}$ | Spin Hall angle when all the spin pumping power is converted to $J_s$ |
| $J_L$ | Charge current flowing at an external load. | $\eta_{out,ab}$ | Conversion efficiency of the absorption power to the output power |
| $P_{out}$ | Output power obtained from an external load. | $J_{c0}$ | Charge current giving rise to emf |
| $P_{out}^{max}$ | Maximum output power that can be acquired from an external load.. | $V_{emf0}$ | Virtual EMF when there is no power consumption |

Figures

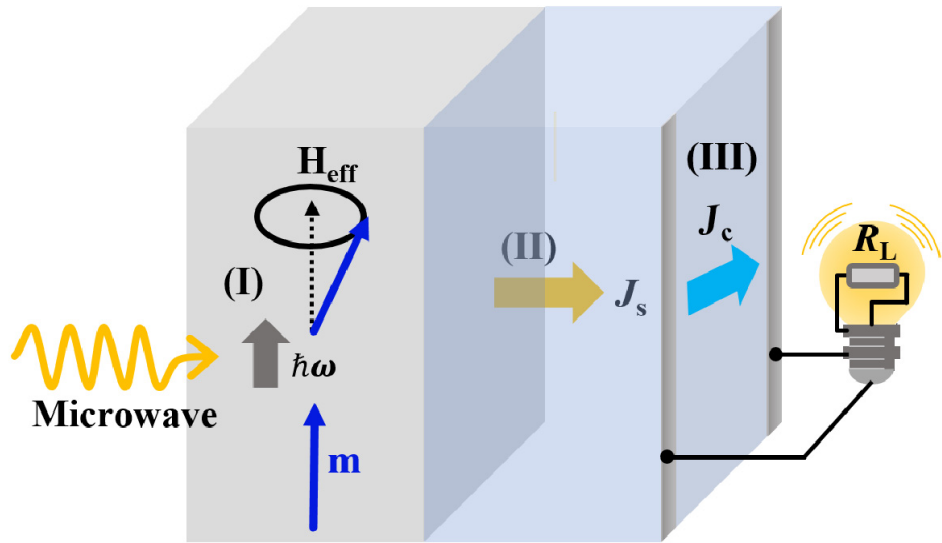

**Fig. 1** Schematic of the overall processes considered in this research. (I) Absorption of microwave in the ferromagnetic (FM) layer, (II) spin pumping from the FM layer into the non-magnetic metal (NM) layer producing spin current ($J_s$) and (III) conversion from $J_s$ to charge current $J_c$ and output of power.

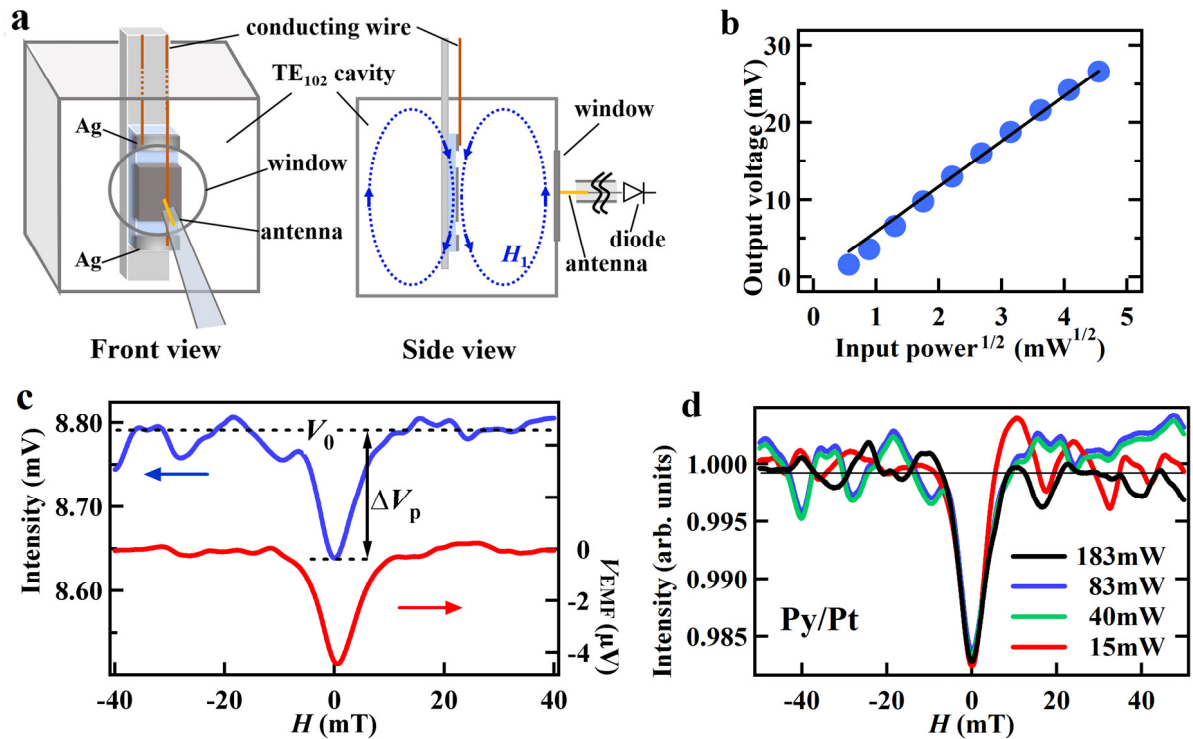

**Fig. 2 a**. Schematic of antenna probe system. The $TE_{102}$ microwave cavity has a window on the front side, from which a little microwave is emitted. The microwave emitted is picked up by an antenna probe set near the window and measured by a detection diode connected to the antenna. **b**. Relation of the output voltage from the microwave detector versus the square root of the input microwave power. The solid line is the result of fit using a linear function. **c**. The absorption spectrum of the permalloy (Py)/Pt bilayer device measured from the antenna probe system (top) and the EMF spectrum measured simultaneously (bottom). $V_0$ is the averaged baseline and $\Delta V_p$ is peak height voltages. **d**. Power dependence of the normalized absorption spectra measured from the antenna system.

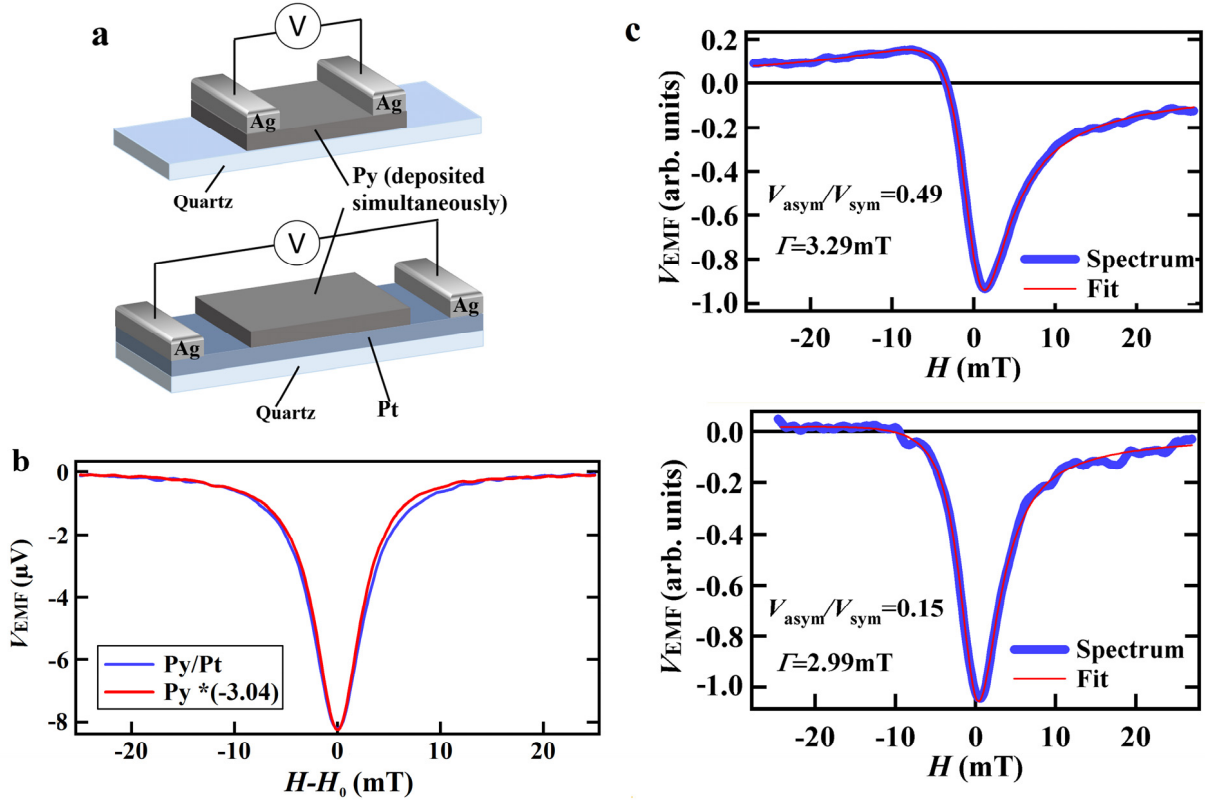

**Fig. 3 a.** Structures of the Py monolayer and Py/Pt bilayer devices used in this study. The Py layers of the two devices were deposited simultaneously by vacuum evaporation techniques. The boundary lines of the Ag electrodes for picking up EMFs in the two devices were fabricated so that the upper and lower electrodes were parallel to each other. **b**. The EMF spectra obtained from the two devices under 162mW microwave irradiation: the intensity of Py device was increased by -3.04 times for comparing the linewidth. **c**. Changes in the EMF spectra of the Py/Pt device due to the mixing of the dispersion spectral component and the results of spectral fit using Eq. (3).

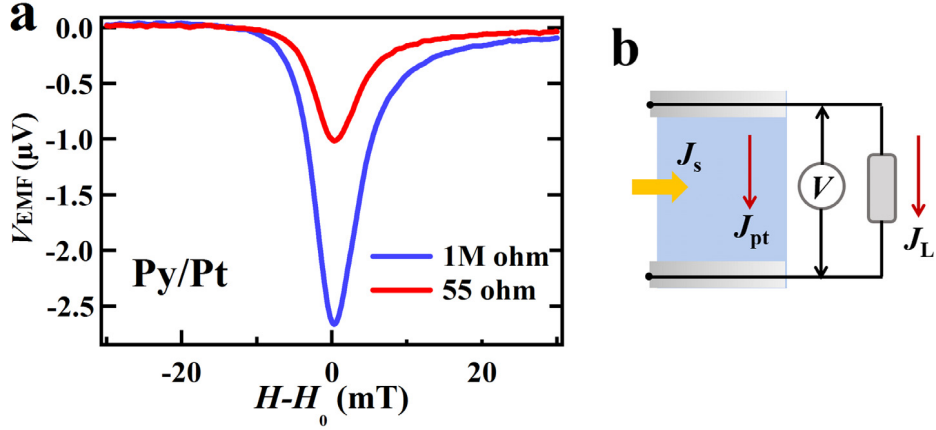

**Fig. 4 a.** Change of EMF spectra depending on the external load resistance ($R_L = 55\Omega$ and $1M\Omega$) under microwave irradiation of 52mW. **b**. Model of the parallel circuit consisting of charge currents flowing inside the Pt layer ($J_{pt}$) and the external load ($J_L$) which are converted from spin currents ($J_s$).

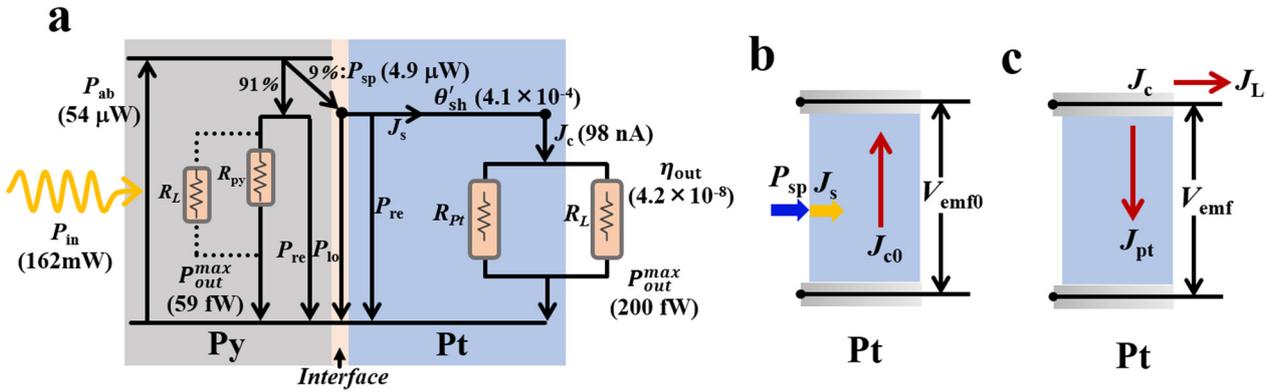

**Fig. 5 Power flow chart in the Py/Pt device. a** Conversion process from the microwave absorption to the power output under 162mW microwave irradiation. $\eta_{out}$ is the conversion efficiency of the power that can be maximally output ($P_{out}^{max}$) per the injected spin pumping power ($P_{sp}$). $\theta'_{sh}$ is the spin Hall angle when all the spin pumping power is converted to $J_s$. $P_{ab}$ is the absorption energy of Py layer. $P_{re}$ and $P_{lo}$ are the powers lost by spin relaxation and by interface loss, respectively. **b** Schematic representation of the power generation process by charging electrodes from the charge current $J_{c0}$ converted from $J_s$. $V_{emf0}$ is the virtual EMF generated by $J_{c0}$ assuming no power consumption. **c** Schematic representation of the actual power consumption process where $J_c$ (= $J_{pt}$+ $J_L$) flows in the opposite direction to $J_{c0}$, by which the observed EMF $V_{emf}$ is much smaller than $V_{emf0}$.